\documentclass{appolbtr}

\usepackage{hyperref}
\usepackage{pstricks,pst-node,pst-text,pst-3d}
\usepackage{colordvi}
\usepackage{epsfig}

\newcommand{\be}{\begin{equation}}
\newcommand{\ee}{\end{equation}}
\newcommand{\bea}{\begin{eqnarray}}
\newcommand{\eea}{\end{eqnarray}}
\newcommand{\bqa}{\begin{eqnarray}}
\newcommand{\eqa}{\end{eqnarray}}

\newcommand{\ep}{\varepsilon}

\begin{document}
  \eqsec  

\title{
{ \begin{flushleft}
DESY 05--215\\
WUE--ITP--2005--013\\
October 2005\\
\vspace*{0.5cm}
\end{flushleft}}
On the massive two-loop corrections to Bhabha scattering%
\thanks{Presented at XXIX Conference of Theoretical Physics:
``Matter to the Deepest'', Ustro\'n, Poland, 8-14 September 2005}%
~\thanks{Work supported in part by
Sonderforschungsbereich/Transregio 9--03 of DFG
``Computergest\"utzte Theoretische Teilchenphysik",  by
the Sofja Kovalevskaja Award of the Alexander von Humboldt Foundation
  sponsored by the German Federal Ministry of Education and Research,
and by the Polish State Committee for Scientific Research (KBN)
for the research project in 2004--2005.
}}

\author{M. Czakon$^{ab}$, J. Gluza$^{bc}$, T. Riemann$^{c}$
\address{$^{a}$ Institut f\"ur Theoretische Physik
und Astrophysik, Universit\"at W\"urzburg\\
Am Hubland, D-97074 W\"urzburg, Germany}
\address{$^{b}$ Institute of Physics, Univ. of
    Silesia, Universytecka 4, 40007 Katowice, Poland}
\address{$^{c}$ Deutsches Elektronen-Synchrotron DESY\\
Platanenallee 6, D--15738 Zeuthen, Germany}
}
\maketitle
\begin{abstract}
We overview the general status of higher order corrections to Bhabha scattering 
and review recent progress in 
the determination of the two-loop virtual corrections.
Quite recently, they were derived from combining a massless calculation and contributions with electron sub-loops.
For a massive calculation, the self-energy and vertex master integrals are known, while most of the two-loop boxes are not.
We demonstrate with an example that a study of
systems of differential equations, combined with
Mellin-Barnes representations for single masters, might open a way for their systematic calculation.
\end{abstract}

\section{Introduction}
Since Bhabha's article \cite{Bhabha:1936xx} on the reaction
\bea
\label{eq1}
e^+e^- \to e^+e^-
\eea
numerous studies of it were performed, triggered by better and better experimentation.
Bhabha scattering is of interest for several reasons.
It is a classical reaction with a clear experimental signature.

Small angle Bhabha scattering has a specific sensitivity to new physics due to its  forward peaking structure, and is a prominent luminosity monitor at high energies.
At LEP, the cross-section predictions were needed with an accuracy of the order of about $10^{-3}$ \cite{Jadach:2003zr}.
At the planned linear $e^+e^-$ collider ILC (see the  documents prepared for TESLA \cite{Aguilar-Saavedra:2001rg}), another order of magnitude ($10^{-4}$) might be needed  \cite{Hawkings:1999ac} and realized \cite{Lohmann:2004nn}.
This accuracy is in reach of the most advanced Bhabha Monte Carlo programs BHLUMI \cite{Jadach:1996is,Jadach:1996md}, NLLBHA \cite{Arbuzov:1995qd,Arbuzov:1996jj}, SAMBHA \cite{Arbuzov:2004wp}.
The programs contain higher order photonic corrections, both exponentiated soft photon corrections and logarithmically enhanced fixed order corrections (including 2nd order).
Concerning the virtual corrections, the logarithmically non-enhanced two-loop contributions (and also radiative one-loop contributions) were not included until recently.

At small energies, large angle Bhabha scattering may be used for luminosity determination.
The KLOE collaboration, e.g., aims at an experimental accuracy of 0.3 \%, compared to a present accuracy estimate of the Monte Carlo programs (BABAYAGA \cite{CarloniCalame:2000pz,CarloniCalame:2003yt}, BHAGENF \cite{Berends:1987jm}, BHWIDE \cite{Jadach:1995nk,Placzek:1999xc}, MCGPJ \cite{Arbuzov:1997je}) of about 0.5 \% \cite{Denig:2005}.
Here, the inclusion of two-loop boxes is not needed.

Many calculations of weak corrections have been performed since the first complete one-loop calculation \cite{Consoli:1979xw}, and the technical accuracy is extremely high nowadays.
A recent one, based on DIANA \cite{Fleischer:2004ah,Gluza:2004tq,Lorca:2004dk} and aItalc \cite{Lorca:2004fg}, demonstrated an agreement of more than 10 digits with another calculation \cite{Gluza:2004tq}.
Leading higher order weak corrections due to the top quark were included in \cite{Bardin:1991xe}, where two calculations were shown to agree within few per mille.
It might also well come out that this accuracy (for the pure weak part) is sufficient for applications in the foreseeable future.
But, it might well be that a two-loop calculation of all the dominant terms of the second order weak corrections is of interest.

The last few years brought considerable progress in the pure photonic, virtual two-loop corrections to Bhabha scattering.
They were determined for the massless case some time ago \cite{Bern:2000ie}, and quite recently the understanding was developed on  how to transform this into the on mass shell renormalization scheme \cite{Penin:2005kf,Penin:2005eh}.
This is an absolutely needed part of the calculation in this approach since the Monte Carlo programs in use rely on a finite electron mass, thus regulating the collinear singularities.
For this, several infra-red properties of the amplitudes had to exploited: exponentiation of the IR logarithms; factorization of the collinear logarithms into external legs; non-renormalization of the IR exponents.
The last of the three properties is not fulfilled for diagrams with an internal fermion loop. For this reason, the massless calculation has to be combined with an explicit calculation of them \cite{Bonciani:2004qt,Bonciani:2003cj,Bonciani:2004gi}.
In  \cite{Bonciani:2005im}, the massive calculation of the Italian/Freiburg group is summarized.
So far, it is
gauge invariant but not complete since the massive box diagrams are an inevitable part of that program.
Further, the present combination of the two existing calculations (Penin's and Italian/Freiburg approach) seems not yet to match exactly what is a physical quantity (Bhabha scattering with several flavors); see also \cite{Bonciani-radcor:2005}.

In view of the importance of the problem and the principal interest it is quite obvious that a complete massive calculation is to be done yet.
A substantial last missing part are the massive two-loop boxes.

\section{Master integrals}
The two-loop vertex master integrals have been completely determined and published by two groups in 2003 \cite{Bonciani:2003te,Bonciani:2003hc} and 2004 \cite{Czakon:2004tg}.
For details and references for the self-energies see \cite{Czakon:2004wm}.
The masters
are available, from both groups, from the internet as Form files or MATHEMATICA files \cite{Czakon:2004n1,Boniani-web:2005}.

Concerning the double boxes, the situation is less developed.
A complete list of masters is known \cite{Czakon:2004wm,Czakon:2004n1}, but the evaluation is, as was expected, quite difficult.
Not only that they depend on a more complicated kinematics compared to vertices or self-energies, namely on both $s$ and $t$.
Additionally one is faced with the fact that up to six  masters (for B6l3m3
\footnote{B6l3m3 is a master topology with six lines, three of them massive; for details of our notations, see \cite{Czakon:2004wm,Czakon:2004n1}.}) may form systems of differential equations, if one wants to go this way.
So it may well happen that the use of Mellin-Barnes representations for the determination of single masters (first successfully used for massless masters \cite{Tausk:1999vh,Smirnov:1999gc}) has to be preferred here. In fact, the heroic efforts of V. Smirnov \cite{Smirnov:2001cm,Heinrich:2004iq} give strong hints to that.
The two-loop Bhabha box diagrams contain Feynman integrals with seven propagators (lines) and with numerators, and their reduction to a smaller number of master integrals leads to the inclusion of masters with a smaller number of lines, and also of masters with dots (with lines raised to a power biger than 1).
The present knowledge of box masters may be summarized shortly:
\begin{itemize}
\item
Seven lines: For topology B7l4m1 (first planar) exist two masters, and B7l4m1 and B7l4m1N are known \cite{Smirnov:2001cm,Heinrich:2004iq}.
For topology B7l4m2 (second planar) we need four masters, and the basic master is known (with use of a two-fold integral in the constant term) \cite{Heinrich:2004iq}.
Finally, for the three masters of topology B7l4m3 (non-planar),  only  the most divergent part ($\sim 1/\ep^2$) of the basic master is determined analytically \cite{Heinrich:2004iq}.
\item
Six lines: The ten masters are not published.
\item
Five lines: Analytical results for singularities are known, see next Section.
\end{itemize}
\section{The box master integrals with five lines}
\begin{figure}
\centering
\includegraphics[width=1\textwidth]{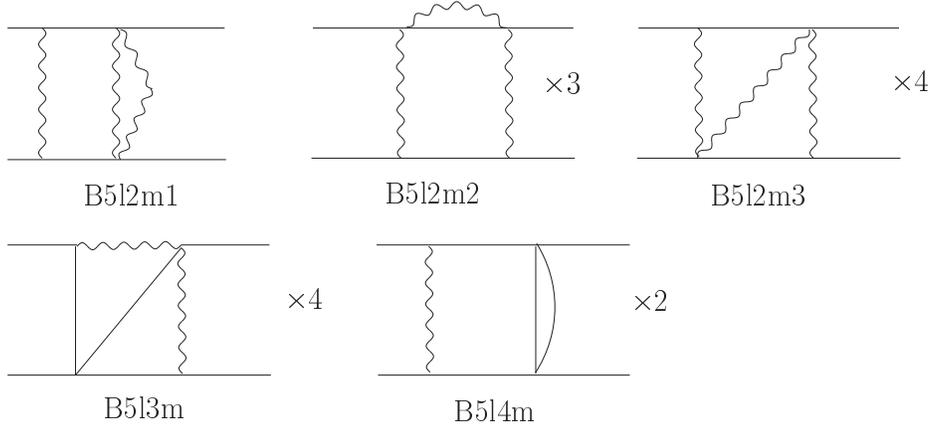}%
\caption{Two-loop MIs with five internal lines}
  \label{five}
\end{figure}

The two-loop box masters depend on two variables $s$ and $t$, or equivalently on related scaling variables $x$ and $y$ (the mass $m = 1$).
There are five master topologies with five internal lines, see Figure \ref{five}.

\begin{figure}
\centering
\includegraphics[width=1\textwidth]{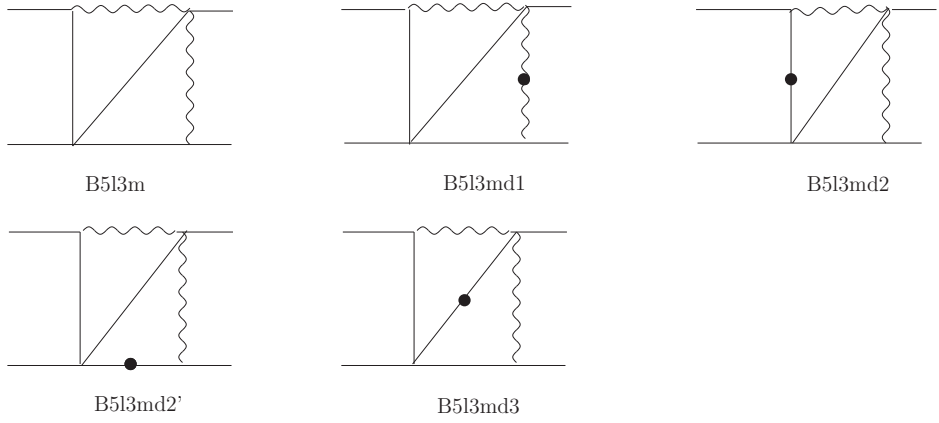}%
\caption{Five MIs for topology B5l3m. Masters B5l3md2' and B5l3md2 are symmetric, B5l2m3d2'[s(x),t(y)]=B5l3md2[t(y),s(x)].}
  \label{b5l3m}
\end{figure}

As experience shows, some of them are already examples of quite complicated systems of master integrals (MIs), although
two cases, namely B5l2m1 (one MI
\cite{ Czakon:2004tg,Czakon:2004n1}) and B5l4m (two MIs
\cite{Bonciani:2003cj,Czakon:2004tg,Czakon:2004n1})
are relatively simple.
B5l2m2 is a system of three MIs, B5l2m3 is one of four MIs, while B5l3m is actually a system of even five MIs (see Fig.~\ref{b5l3m}).
In \cite{Czakon:2004wm}, analytical results for the singularities of topologies B5l2m2 and B5l2m3 have been given.
Here we complete the determination of singularities of the MIs with five lines and consider the most complex case B5l3m.
\subsection{B5l3m: Synergetic use of Mellin-Barnes representations and differential equations}
If we try to calculate the masters for topology B5l3m with differential equations (DEs), we meet a system of five
coupled differential equations.
In fact, one of the dotted masters (B5l2m3d2) appears twice (with interchanged arguments).

Fortunately, we can use another method, based on 
Mellin-Barnes presentations, to solve some of them separately. We found it most efficient to solve
the case B5l3md2, where one eliminates two masters at once.
Details of the calculation will be given elsewhere \cite{Riemann-radcor:2005}.
The result is:
\begin{eqnarray}
{\tt B5l3md2} &=& - \frac{1}{\epsilon^2}
\frac{x}{4 (1 - x^2)} H[0, x] \nonumber \\
&+&
\frac{1}{\epsilon} \frac{x}{4 (1 - x^2) (1 - y^2)}  \left[ 2 (1 + y^2) H[0, x]
H[0, y] \right. \nonumber \\
&-& \left. (-1 + y^2) (\xi_2 + 6 H[-1, 0, x] - 4 H[0, 0, x] -
               2 H[1, 0, x]) \right] \nonumber \\
& +& {\cal{O}}(1),
\label{d2}
\end{eqnarray}
Then, up to (and including) order $1/\epsilon$, we are left with a system of only three DEs.
The differential equation for B5l3md3[-1] is:
\bqa
\frac{\partial  {B5l3md3[-1]}}{\partial x} &=&
\frac{1+x^2}{x(1-x^2)}{B5l3md3[-1]} - \frac{yH[0,y]}{(1-x^2)(1-y^2)},
\eqa
with $ s=-(1-x)^2/x$, $ t=-(1-y)^2/y$, and  can also be solved separately.
The solution is:
\bqa
{B5l3md3[-1]} = -\frac{xy}{(-1 + x^2)(-1 + y^2)} {H[0, x]} {H[0, y]}.
\label{d3}
\eqa
For our notations in terms of HPLs and generalized HPLs see \cite{Czakon:2005jd} and references therein.

Finally we have to determine the singularities of B5l3md1. They can be found from algebraic manipulations (see \cite{Czakon:2004wu}).
This relies on the fact
that the basic integral, B5l3m, and the integrals with numerators are nonsingular.
In order to use this, we replaced in the defining equations for B5l3md1 the dotted diagrams by those with numerators.
The solution is:
\begin{eqnarray}
{\tt B5l3md1} &=& \frac{1}{\epsilon^2} \frac{1}{8 x (-1 + y) (1 + y)^3} \nonumber \\
&\times & \left[ (-1 + x)^2 y (-1 + y^2 + 2 y H[0, y]) \right] \nonumber \\
&+&
\frac{1}{\epsilon} \frac{y}{24 x (1 + x) (-1 + y) (1 + y)^3)} \nonumber \\
&\times & \left[ 6 (-1 + x - x^2 + x^3) H[0, x] 
  (-1 + y^2 + 2 y H[0, y]) \right. \nonumber \\
&-& 6 (1 + x) (-2 - 2 x^2 + 2 y^2 + 2 x^2 y^2 +
  y \xi_2 - 2 x y \xi_2 \nonumber \\ 
&+& x^2 y \xi_2 + 2 (-2 x - y + 2 x y - x^2 y - 2 x y^2 \nonumber \\
&+&
  (-1 + x)^2 y H[-1, -y] + 3 (-1 + x)^2 y H[-1, y]) H[0, y] \nonumber \\
&-&
  6 (-1 + x)^2 y H[0, -1, y] -4 y H[0, 0, y] + 8 x y H[0, 0, y] \nonumber \\
&-& \left.
  4 x^2 y H[0, 0, y] + 2 y H[0, 1, y] - 4 x y H[0, 1, y] +
  2 x^2 y H[0, 1, y]) \right] \nonumber \\
& +& {\cal{O}}(1),\label{d1}
\end{eqnarray}
We can easily check that the solutions (\ref{d1}) and (\ref{d2}) fulfill the original differential equations (not reproduced here for space reasons).
The knowledge of the solution
(\ref{d2}), obtained by the Mellin-Barnes method, has been used.
So, our combined approach is also a cross check for the analytical
result obtained with the Mellin-Barnes method.
Finally, we checked that (\ref{d1}), (\ref{d3}), and (\ref{d2}) are also in agreement with a numerical approach based
on sector decomposition calculations.

\section{Summary}
We gave an overview on status and recent progress for the higher order corrections for Bhabha scattering.
The small angle Monte Carlo programs are approaching the accuracy standard of 10$^{-4}$ once this is needed.

The results of the determination of the pure constant (non-logarithmically enhanced) virtual two-loop corrections seem to stay perhaps negligibly small (at this accuracy level and after isolating from them not only the mass-, but also the kinematically enhanced terms) in the kinematical region of interest, but this is not finally numerically fixed.

In order to have a complete massive calculation (though in the limit of small electron mass), one has to determine yet the bulk of two-loop boxes.
This is underway.
\providecommand{\href}[2]{#2}


\begin{thebibliography}{10}

\bibitem{Bhabha:1936xx}
H. Bhabha,
\newblock Proc. Roy. Soc. A154 (1936) 195.

\bibitem{Jadach:2003zr}
S. Jadach,
\newblock hep-ph/0306083.

\bibitem{Aguilar-Saavedra:2001rg}
ECFA/DESY LC Physics Working Group, J. Aguilar-Saavedra et~al.,
\newblock (2001), hep-ph/0106315,
\newblock TESLA TDR, Part III: Physics at an $e^+e^-$ Linear Collider, DESY
  2001-011.

\bibitem{Hawkings:1999ac}
R. Hawkings and K. M{\"o}nig,
\newblock Eur. Phys. J. direct C1 (1999) 8, hep-ex/9910022.

\bibitem{Lohmann:2004nn}
H. Abramowicz et~al.,
\newblock IEEE Transactions on Nuclear Science 51 (2004) 1.

\bibitem{Jadach:1996is}
S. Jadach et~al.,
\newblock Comput. Phys. Commun. 102 (1997) 229.

\bibitem{Jadach:1996md}
S. Jadach et~al.,
\newblock Nucl. Phys. Proc. Suppl. 51C (1996) 164, hep-ph/9607358.

\bibitem{Arbuzov:1995qd}
A. Arbuzov et~al.,
\newblock Nucl. Phys. B485 (1997) 457, hep-ph/9512344.

\bibitem{Arbuzov:1996jj}
A. Arbuzov et~al.,
\newblock Nucl. Phys. Proc. Suppl. 51C (1996) 154, hep-ph/9607228.

\bibitem{Arbuzov:2004wp}
A. Arbuzov et~al.,
\newblock Eur. Phys. J. C34 (2004) 267, hep-ph/0402211.

\bibitem{CarloniCalame:2000pz}
C.M. Carloni~Calame et~al.,
\newblock Nucl. Phys. B584 (2000) 459, hep-ph/0003268.

\bibitem{CarloniCalame:2003yt}
C.M. Carloni~Calame et~al.,
\newblock Nucl. Phys. Proc. Suppl. 131 (2004) 48, hep-ph/0312014.

\bibitem{Berends:1987jm}
F.A. Berends, R. Kleiss and W. Hollik,
\newblock Nucl. Phys. B304 (1988) 712.

\bibitem{Jadach:1995nk}
S. Jadach, W. Placzek and B.F.L. Ward,
\newblock Phys. Lett. B390 (1997) 298, hep-ph/9608412.

\bibitem{Placzek:1999xc}
W. Placzek et~al.,
\newblock hep-ph/9903381.

\bibitem{Arbuzov:1997je}
A.B. Arbuzov et~al.,
\newblock JHEP 10 (1997) 006, hep-ph/9703456.

\bibitem{Denig:2005}
A. Denig,
\newblock Bhabha Scattering at DAFNE: The KLOE Luminosity Measurement, talk
  held at Bhabha Mini-Workshop, Karlsruhe, 21-22 April 2005.

\bibitem{Consoli:1979xw}
M. Consoli,
\newblock Nucl. Phys. B160 (1979) 208.

\bibitem{Fleischer:2004ah}
J. Fleischer, A. Lorca and T. Riemann,
\newblock hep-ph/0409034.

\bibitem{Gluza:2004tq}
J. Gluza, A. Lorca and T. Riemann,
\newblock Nucl. Instrum. Meth. A534 (2004) 289, hep-ph/0409011.

\bibitem{Lorca:2004dk}
A. Lorca and T. Riemann,
\newblock Nucl. Phys. Proc. Suppl. 135 (2004) 328, hep-ph/0407149.

\bibitem{Lorca:2004fg}
A. Lorca and T. Riemann,
\newblock hep-ph/0412047.

\bibitem{Bardin:1991xe}
D. Bardin, W. Hollik and T. Riemann,
\newblock Z. Phys. C49 (1991) 485.

\bibitem{Bern:2000ie}
Z. Bern, L. Dixon and A. Ghinculov,
\newblock Phys. Rev. D63 (2001) 053007, hep-ph/0010075.

\bibitem{Penin:2005kf}
A.A. Penin,
\newblock  hep-ph/0501120.

\bibitem{Penin:2005eh}
A.A. Penin,
\newblock hep-ph/0508127.

\bibitem{Bonciani:2004qt}
R. Bonciani et~al.,
\newblock Nucl. Phys. B716 (2005) 280, hep-ph/0411321.

\bibitem{Bonciani:2003cj}
R. Bonciani et~al.,
\newblock Nucl. Phys. B681 (2004) 261, hep-ph/0310333.

\bibitem{Bonciani:2004gi}
R. Bonciani et~al.,
\newblock Nucl. Phys. B701 (2004) 121, hep-ph/0405275.

\bibitem{Bonciani:2005im}
R. Bonciani and A. Ferroglia,
\newblock Phys. Rev. D72 (2005) 056004, hep-ph/0507047.

\bibitem{Bonciani-radcor:2005}
R. Bonciani,
\newblock Two-Loop Bhabha Scattering in QED, talk held at RADCOR 2005,\\ Shonan
  Village, Japan, 2-7 October 2005, to appear in the proceedings.

\bibitem{Bonciani:2003te}
R. Bonciani, P. Mastrolia and E. Remiddi,
\newblock Nucl. Phys. B661 (2003) 289, hep-ph/0301170.

\bibitem{Bonciani:2003hc}
R. Bonciani, P. Mastrolia and E. Remiddi,
\newblock Nucl. Phys. B690 (2004) 138, hep-ph/0311145.

\bibitem{Czakon:2004tg}
M. Czakon, J. Gluza and T. Riemann,
\newblock Nucl. Phys. (Proc. Suppl.) B135 (2004) 83, hep-ph/0406203.

\bibitem{Czakon:2004wm}
M. Czakon, J. Gluza and T. Riemann,
\newblock Phys. Rev. D71 (2005) 073009, hep-ph/0412164.

\bibitem{Czakon:2004n1}
M. Czakon, J. Gluza and T. Riemann,
\newblock DESY Zeuthen homepage on 2-loop {Bhabha} scattering,
  http://www-zeuthen.\linebreak[2]desy.\linebreak[2]de/%
  \linebreak[2]theory/\linebreak[2]research\linebreak[2]/bhabha/bhabha.html.

\bibitem{Boniani-web:2005}
R. Bonciani,
\newblock Webpage on Bhabha scattering,\\
  http://pheno.physik.uni-freiburg.de/${\sim}$bhabha/.

\bibitem{Tausk:1999vh}
J.B. Tausk,
\newblock Phys. Lett. B469 (1999) 225, hep-ph/9909506.

\bibitem{Smirnov:1999gc}
V.A. Smirnov,
\newblock Phys. Lett. B460 (1999) 397, hep-ph/9905323.

\bibitem{Smirnov:2001cm}
V. Smirnov,
\newblock Phys. Lett. B524 (2002) 129, hep-ph/0111160.

\bibitem{Heinrich:2004iq}
G. Heinrich and V. Smirnov,
\newblock Phys. Lett. B598 (2004) 55, hep-ph/0406053.

\bibitem{Riemann-radcor:2005}
T.R. M.~Czakon, J.~Gluza,
\newblock Massive Two-Loop Bhabha Scattering, talk held at RADCOR 2005,\\
  Shonan Village, Japan, 2-7 October 2005, to appear in the proceedings.

\bibitem{Czakon:2005jd}
M. Czakon, J. Gluza and T. Riemann,
\newblock  hep-ph/0508212.

\bibitem{Czakon:2004wu}
M. Czakon, J. Gluza and T. Riemann,
\newblock On master integrals for two loop {Bhabha} scattering, hep-ph/0409017.

\end{thebibliography}
\end{document}